\documentclass{article}

\usepackage{PRIMEarxiv}

\usepackage[utf8]{inputenc} % allow utf-8 input
\usepackage[T1]{fontenc}    % use 8-bit T1 fonts
\usepackage{hyperref}       % hyperlinks
\usepackage{url}            % simple URL typesetting
\usepackage{booktabs}       % professional-quality tables
\usepackage{amsfonts}       % blackboard math symbols
\usepackage{nicefrac}       % compact symbols for 1/2, etc.
\usepackage{microtype}      % microtypography
\usepackage{fancyhdr}       % header
\usepackage{graphicx}       % graphics
\usepackage{float}
\usepackage{subfigure}

\usepackage{graphicx}%
\usepackage{multirow}%
\usepackage{amsmath,amssymb}%
\usepackage{amsthm}%
\usepackage{mathrsfs}%
\usepackage[title]{appendix}%
\usepackage{xcolor}%
\usepackage{textcomp}%
\usepackage{manyfoot}%
\usepackage{algorithm2e}
\title{Warped multifidelity Gaussian processes for data fusion of skewed environmental data
}

\author{
  Pietro Colombo \\
  School of Mathematics and Statistics \\
  University of Glasgow, UK \\
  \texttt{Pietro.Colombo@glasgow.ac.uk} \\
  \And
  Claire Miller \\
  School of Mathematics and Statistics \\
  University of Glasgow, UK \\
  \texttt{Claire.Miller@glasgow.ac.uk} \\
  \And
  Xiaochen Yang \\
  School of Mathematics and Statistics \\
  University of Glasgow, UK \\
  \texttt{Xiaochen.Yang@glasgow.ac.uk} \\
  \And
  Ruth O'Donnell \\
  School of Mathematics and Statistics \\
  University of Glasgow, UK \\
  \texttt{Ruth.Haggarty@glasgow.ac.uk} \\
  \And
  Paolo Maranzano \\
  Department Economics, Management and Statistics (DEMS) \\ 
  University of Milano-Bicocca, Italy \& \\
  Fondazione Eni Enrico Mattei (FEEM), Milano, Italy \\
  \texttt{paolo.maranzano@unimib.it} \\
}

\begin{document}
\maketitle

\begin{abstract}
Understanding the dynamics of climate variables is paramount for numerous sectors, like energy and environmental monitoring. This study focuses on the critical need for a precise mapping of environmental variables for national or regional monitoring networks, a task notably challenging when dealing with skewed data. To address this issue, we propose a novel data fusion approach, the \textit{warped multifidelity Gaussian process} (WMFGP). The method performs prediction using multiple time-series, accommodating varying reliability and resolutions and effectively handling skewness. In an extended simulation experiment the benefits and the limitations of the methods are explored, while as a case study, we focused on the wind speed monitored by the network of ARPA Lombardia, one of the regional environmental agencies operting in Italy. ARPA grapples with data gaps, and due to the connection between wind speed and air quality, it struggles with an effective air quality management. We illustrate the efficacy of our approach in filling the wind speed data gaps through two extensive simulation experiments. The case study provides more informative wind speed predictions crucial for predicting air pollutant concentrations, enhancing network maintenance, and advancing understanding of relevant meteorological and climatic phenomena.
\end{abstract}

% keywords can be removed
\keywords{Data Fusion \and Multifidelity Gaussian process \and Skew Data \and ARPA Lombardia}

\section{1. Introduction}
Gaining insight into the changes in environmental variables like wind speed is crucial, especially in fields such as the energy sector, where it influences decisions about wind farm development, and in environmental monitoring to assess air quality in specific regions. These tasks range from mitigating atmospheric and acoustic pollution to overseeing water quality and electromagnetic fields, reflecting the broader mission of environmental stewardship and protection, see \cite{maranzano2022air}. Data fusion algorithms are becoming highly successful in supporting these challenges, as they intuitively combine multiple data sources to obtain more informative data and hence a potentially more cost-effective results. Our study extends the class of data fusion algorithms based on Gaussian processes, referred to as multifidelity, specifically by incorporating a non-parametric warping function for response variables of interest. This approach allows us to effectively merge data having various reliability (fidelity) from different monitoring stations, even when dealing with skewed data distributions, a common occurrence in environmental variables such as wind speed, temperature, and precipitation. More specifically, we propose an extension of the autoregressive multifidelity Gaussian process (MFGP) as proposed by \cite{le2015cokriging} and used the non-parametric warping of \cite{agou2022spatial} to handle skewed data. Our extension, called \textit{warped multifidelity Gaussian process} (WMFGP), maintains the same assumptions of the standard MFGP method, where it is assumed a highly reliable source of information, labelled high-fidelity (HF) is related to a source of information, namely low-fidelity (LF), regarded as less reliable, connected together by a so called autoregressive representation of fidelity levels through a regressive parameter \textit{$\rho$}. This implies that the effectiveness of data fusion of the low-fidelity with high-fidelity data is directly influenced by the linear correlation between the two data sources and more precisely a higher correlation leads to more effective recovery of high-fidelity data.

We apply our model to wind speed data provided by one of the regional environmental protection agencies (in Italian, \textit{Agenzia Regionale di Protezione dell'Ambiente}) active in Italy, namely, ARPA Lombardia, which operates in the Northern part of the country. 
The Lombardy region, nestled in the Po River valley and surrounded by the Alps, faces unique challenges in air quality management. In fact, its geographical configuration limits air circulation, contributing to elevated pollutant concentrations \cite{maranzano2022air}. Furthermore, Lombardy ranks among Europe's most industrialized regions (see Eurostat database \cite{eurostat}), hosting mechanical, electrical, metallurgic, textile, and chemical industries, along with numerous animal farms \cite{colombo2023assessing}, which recent studies have linked to significant impacts on air quality \cite{fasso2023agrimonia}. There are multiple studies where it is demonstrated that wind speed plays a critical role in determining air pollutant dispersion, such as those presented in \cite{RN415}, \cite{RN416}, \cite{RN417}, \cite{RN418} and \cite{RN519}. Hence, an accurate mapping of wind speed across the region is essential for understanding and managing air quality effectively. 

ARPA Lombardia manages a broad weather and air quality monitoring network, which sometimes exhibit data gaps, not only in air pollutant concentration but also in wind speed data. Filling these gaps is vital for comprehensively understanding wind patterns and their implications for air quality. For example, ARPA could schedule the maintenance of the monitoring stations more effectively, if accurate methods for filling data gaps were available. Moreover, data gaps might occur in the presence of particular meteorological events. The absence of a week's worth of data might result in a lack of understanding of the physical generation process.
The remainder of the paper is organized as follows: introduction to the methodology, including a historical overview of multifidelity methods and the challenges associated with skewness. Two experiments are described to demonstrate WMFGP's efficacy in handling skewness in a data fusion context. 
The practical application is described and the results presented.

\section{2. Methods} \label{T & M}

\subsection{2.1 Historical overview}
In the literature the word \textit{fidelity} is used in place of the word quality or reliability (\cite{MF_MassiveDatasets}, \cite{perdikaris2017nonlinear}, \cite{le2015cokriging}). The first appearance of a multifidelity method in the form of a Gaussian process (GP) can be traced back to 2000, when Kennedy and O'Hagan introduced the idea of creating a model to operate at different fidelity levels in their work \cite{kennedy2000predicting}. The intuition was that such a model could levarage the qualities of different datasets. Indeed, HF data are usually scarce in time and space and very expensive to collect. While, LF are typically cheaper and more abundant in time and space, but unreliable by definition. More recently, the framework elaborated by Kennedy and A'Hagan is more commonly referred to as multifidelity. The approach is highly flexible as it can manage data of different reliability and resolutions. Indeed, most of the institutional ground monitoring networks, such as the one managed by the Italian ARPAs, suffer from a poor spatial coverage and the presence of missing sequences representing a optimal case study for multifidelity methods. Our application deals with the this latter problem, using multifidelity methods to recover missing sequences.

Research on multifidelity models has investigated multiple aspects. For example, the paper by \cite{le2015cokriging} focused on the estimation procedure, recasting the equations of the original inefficient Bayesian approach, proposing a sequential design and enabling a much faster inference. Some researchers instead modelled the relationship between the HF and the LF data, as seen in \cite{perdikaris2017nonlinear}, where the authors propose a multifidelity model to handle nonlinear relationships between fidelity levels, since the original model assumed a linear relationship. Instead in \cite{raissi2016deep} the authors combined the multifidelity Gaussian process with a neural network to create a robust model for discontinuous relationships. An excellent overview of different multifidelity methods can be found in \cite{costabal2019multi}. 

The autoregressive multifidelity model, as described in \cite{kennedy2000predicting}, \cite{le2015cokriging}, and \cite{perdikaris2017nonlinear}, assumes the existence of different Gaussian processes modelling the various fidelity levels. Therefore, like the standard Gaussian process, the MFGP is based on normality assumptions, specifically in the error term. However, environmental data such as wind speed, relative humidity, temperature, and precipitation often follow skewed distributions, which renders the normality assumption inappropriate (\cite{aslam2021study}). Addressing non-normal properties using a Gaussian process has historically been a challenge. One intuitive approach might involve using a multivariate skew-normal distribution class. However, as discussed by \cite{genton2012identifiability}, such distribution classes suffer from identifiability issues in inference. For instance, the sum of i.i.d. skew normal random variables is not a skew normal random variable.
Many attempts have been made to mitigate these issues. For example, \cite{alodat2020gaussian} proposed a Gaussian process with skew errors based on a closed skew normal distribution. However, the model proved to be heavily parameterized, making inference computationally challenging. A recent alternative has been formalized in \cite{khaledi2023spatial}, which proposes a parsimonious version of the closed skew normal distribution for use within a Gaussian process framework. This offers an intriguing approach to test in real-case scenarios. A completely different approach from the one of Khaledi or Alodat and Shakhatreh for addressing non-normal properties, particularly skewness, involves transformations. The idea is to transform the response variable in such a way that it follows a normal distribution. These transformations can be either parametric or non-parametric. An example of a parametric transformation is the Box-Cox transformation. However, parametric transformations like the Box-Cox rely on parameters that are data-dependent.  In a multifidelity context, where multiple data sources need to be transformed, the optimal parameters required for appropriate normalization of each data source differ. Applying different transformations might alter the relationship between the HF and LF data. Within the class of parametric transformations for Gaussian processes  there is also the warped Gaussian process \cite{snelson2003warped}, in which a non-linear mapping of the response variable is directly incorporated into the likelihood function such that the GP parameters are learned jointly with the parameters of the normalizing function. This classic approach is  very well suited for multiple application. However, sometimes the transformations learned in this way are not capable of finding effective normalization. A non-parametric non-linear transformation of the response variable for GP is proposed in \cite{agou2022spatial}. The data driven transformation learned in such a way, also referred to as warping, is much more flexible than parametric approaches since it can effectively normalize any data source as long as enough data are provided. More importantly, since the proposed method's transformation is based on the quantiles of a CDF, the normalized values preserve the same ordering in normalized space (latent space). This property is also referred as \textit{quantile invariance}, and ensure that the relationships between different datasets does not change in the latent space. %Finally, it is essential to acknowledge that deciding to employ multifidelity methods or stick with \textit{single-fidelity} methods involves a trade-off, a well-studied topic in the literature. Yet, this particular aspect falls outside the scope of our paper, where our primary focus is on an effective skewness treatment. Moreover, the decision to use multifidelity methods is highly application-dependent, and researchers should evaluate its suitability for their specific use cases.

\subsection{2.2 The autoregressive multifidelity Gaussian process}\label{Section MFGP}

We will first describe the autoregressive multifidelity model, as our method adopts an extension of this. Let's consider the presence of a set of observations denoted as $\mathbf{y_{H}}$, which are independent and identically distributed (i.i.d) and span from indices 1 to $N_H$. These observations are collected at various locations denoted as $\mathbf{x_{H}}$. We also refer to the observed response at any given location as $y(\mathbf{x_{H}})$. The subscript $\mathbf{H}$ is used to emphasize that these observations are sourced from high fidelity data, implying that if they pertain to an environmental factor such as wind speed, the measured wind speed closely approximates the actual, unobserved wind speed, denoted as $u_H(\mathbf{x_{H}})$. However, it is worth noting that these observations, $y(\mathbf{x_{H}})$, are often limited in both spatial and temporal coverage. Constructing a precise model becomes especially challenging when the sample size $N_H$ is small. Nevertheless, there may be other observations that are correlated with the variable of interest and are available near the spatial location of $y(\mathbf{x_{H}})$. For instance, these additional lower fidelity data sources could originate from other monitoring stations. Let us call these observations $\mathbf{y_L}$, and they are also i.i.d, but observed at location $\mathbf{x_L}$, of sample size $N_L>>N_H$. Moreover, we assume that $\mathbf{x_H} \subset \mathbf{x_L}$ meaning that we have a nested sampling design, in which each individual $i$ of high fidelity observation $\mathbf{y^i_H}$  has a  respective $\mathbf{y^i_L}$, but not vice-versa. The latter assumption is fundamental as it permits the application of the recursive estimation schemes elaborated in \cite{le2015cokriging}. The observations $y(\mathbf{x_L})$ are measured with lowered degree of reliability. Such a situation can be modelled by means of Gaussian process regression (GPR); see \cite{williams2006gaussian} for more details.
Then, we could formalize the following model :
\begin{equation}
   y(\mathbf{x_L}) = u_L(\mathbf{x_L}) + \mathbf{\epsilon_L(x_L)}, \ \ \ \ \mathbf{\epsilon_L} \stackrel{\text{i.i.d}}\sim \mathcal{N}(\mathbf{0}, \Sigma_L^2 \mathbf{I}). 
\end{equation}
In such an equation, $u_L(\mathbf{x_L})$ represents the true unobserved phenomena of the LF data, and it is a $\mathcal{GP}(0, C(x',x, \theta_1))$ that depends on the generic covariance function $C(x',x, \theta_1)$, characterized by the vector of parameters $\theta_1$. Note that the error term $\mathbf{\epsilon_L}$ is an i.i.d and normally distributed error and describes the discrepancy between the observed responses $y(\mathbf{x_L})$ and the true low fidelity process $u_L(\mathbf{x_L})$. While, the true process underlining the high quality is related to the LF process by the following recursive equation: 
% \textcolor{red}{riscrivere equazione come sistema o togliere l'errore. Ho parli di osservati o di processi, i processi sono senza erroree.}
\begin{equation}\label{Eq2}
    u_H(\mathbf{x_L})= \rho u_L(\mathbf{x_L}) + \delta(\mathbf{x_L})+\epsilon_{\delta}(\mathbf{x_L}). 
\end{equation}

The latter equation establishes a relationship between a high-quality function $u_H(\cdot)$ and a low-quality function $u_L(\cdot)$ at each location at which the low-quality information is available $\mathbf{x_L}$. The function $\delta(\cdot)$ is an independent $\mathcal{GP} \sim (0,C(x',x, \theta_2)$
that is sometimes referred to as the discrepancy process since it describes the discrepancy between HF and LF data, while $\rho$ is the scaling constant that quantifies the relation between the two datasets. The measurement error of the discrepancies is expressed by $\epsilon_{\delta} \stackrel{\text{i.i.d}} \sim N(0,\sigma_{\delta}^2)$, such a noise might be related to the fact the high fidelity data are measured with some error that could affect the discrepancies derivation. The latter assumption, is an assumption of normality that renders the model inadequate to deal with skew discrepancies. Equation \ref{Eq2} intuitively defines also the name of the model as the $\rho$ sometimes is called as autoregressive parameter, given that it establishes an autoregressive relationship between fidelity levels, where $u_L(\cdot)$
is labelled as fidelity $t-1$, and $u_H(\cdot)$ as fidelity $t$. In other words, the term autoregressive does not refer to any ordering in time or space but to the degree of fidelity. The model can be generalized for multiple fidelity levels.
From a practical point of view, the full model can be thought of as a standard Gaussian process where the vector of observations is:
\begin{equation}
    \mathbf{y} = \begin{bmatrix}
\mathbf{y}_L \\
\mathbf{y}_H
\end{bmatrix}
\end{equation}
while the covariance matrix is defined as follows:

\begin{equation}    
% \tiny
\mathbf{K} = \begin{bmatrix}
C_{LL}(\mathbf{x}_L, \mathbf{x}_L; \theta_1) + \sigma_L^2 \mathbf{I} & C_{LH}(\mathbf{x}_L, \mathbf{x}_H; \theta_1, \rho)\\
C_{HL}(\mathbf{x}_H, \mathbf{x}_L; \theta_1) & C_{HH}(\mathbf{x}_H, \mathbf{x}_H; \theta_1, \theta_2, \rho) + \sigma_H^2 \mathbf{I}
\end{bmatrix}. 
% \normalsize
\end{equation}
The $\mathbf{K}$ matrix depends on multiple covariance functions defined by the positions of  $\mathbf{x_L}$,  of $\mathbf{x_L}$ and $\mathbf{x_H}$,
jointly, or on only by $\mathbf{x_H}$. Moreover, the 
term $C_{HH}(\mathbf{x}_H, \mathbf{x}_H; \theta_1, \theta_2, \rho) + \sigma_H^2 \mathbf{I}$, which defines the correlation between the high-fidelity data, is designed using the parameters of the discrepancy process and the low-fidelity process. More precisely:

\begin{equation}
  C_{HH}(\mathbf{x_H},\mathbf{x_H};\theta_1,\theta_2,\rho) = \rho^2 C(\mathbf{x_H},\mathbf{x_H};\theta_1) + C(\mathbf{x_H},\mathbf{x_H};\theta_2).
\end{equation}
The model for two fidelity levels has seven parameters, doubling the standard Gaussian process parameters, the nugget (here inside the vector $\theta$), the signal variance $\sigma_H$ and $\sigma_L$, and length scale (inside the vector $\theta$ as well) plus the autoregressive parameter $\rho$.
The estimation can be derived by minimizing the negative log-likelihood:
\begin{equation}
    \mathcal{NLML}(\theta_1, \theta_2, \rho) = \frac12 \mathbf{y}^T \mathbf{K}^{-1} \mathbf{y} + \frac12 \log |\mathbf{K}| + \frac{N}{2} \log(2 \pi),
\end{equation}
and for new input locations $\mathbf{x}^*$, the predictions are returned by the following mean and variance equations:

\begin{align} \label{eq7} 
   u_H(\mathbf{x^*})=\mathbf{q}^T \mathbf{K}^{-1} \mathbf{y}\\ \label{eq8}
 V(u_H(\mathbf{x^*}))  =C_{HH}(\mathbf{x}^*,\mathbf{x}^*) - \mathbf{q}^T\mathbf{K}^{-1}\mathbf{q} 
\end{align}
with $\mathbf{q}$ being equal to $\mathbf{q}^T = \begin{bmatrix}
C_{HL}(\mathbf{x}^*,\mathbf{x}_L; \theta_1, \rho) & C_{HH}(\mathbf{x}^*,\mathbf{x}_H; \theta_1, \theta_2, \rho)
\end{bmatrix}.$

\subsection{2.4 Warped multifidelity Gaussian process}

The warped multifidelity approach involves a nonlinear and non-parametric transformation of the non-normal data $\mathbf{y_H}$ and $\mathbf{y_L}$ into a latent space where they are normalized. Subsequently, a multifidelity model is run in this latent space. We refer to the normalized data as $\mathbf{Ny}_H$ and $\mathbf{Ny}_L$. The two nonlinear mappings, $g_L: \mathbf{y_L} \rightarrow \mathbf{Ny}_L$ and $g_H: \mathbf{y_H} \rightarrow \mathbf{Ny}_H$, are designed such that $Ny(\mathbf{x_L}) = g_L[y(\mathbf{x_L})]$ and $Ny(\mathbf{x_H}) = g_H[y(\mathbf{x_H})]$ exhibit both marginal and joint normal distributions. The warping transformations, $g_L(\cdot)$ and $g_H(\cdot)$, are obtained by first computing CDF$_H$ and its marginal $F_{y}(y_H)$ for the high-fidelity data, and CDF$_L$ and its marginal $F_{y}(y_L)$ for the low-fidelity data. Then, normal scores are obtained by inverting these cumulative distribution functions (CDFs). More precisely, with $\phi$ being a normal CDF we have:

\begin{align}\label{eq10 and 11}
\mathbf{Ny}_L &= \phi_L(F{y}(\mathbf{y_L}))^{-1} = g_L(\mathbf{y_L}) \\
\mathbf{Ny}_H &= \phi_H(F{y}(\mathbf{y_H}))^{-1} = g_H(\mathbf{y_H}).
\end{align}
This means the procedure needs to estimate the CDF from the data. The estimation of the CDF, in turn, is obtained by integrating an antecedent kernel density; more details of the specific kernel density estimation are provided in \cite{pavlides2022non}. Once the multifidelity model is trained using the latent normalized observations $\mathbf{Ny}_L$ and $\mathbf{Ny}_H$, it becomes necessary to back-transform the estimated quantities (the means and variances of the multifidelity model, as shown in equations \ref{eq7} and \ref{eq8}) into the original space. Therefore, we need an inverse warping transform, referred to as $\Tilde{g}(\cdot) = g(\cdot)^{-1}$. In other words, $\Tilde{g}(\cdot):\mathbf{Ny} \rightarrow \mathbf{y}.$ This latter inverse transformation is also a monotonic mapping and it is obtained by means of a lookup table, see \cite{agou2022spatial}, since an explicit version of the inverse function is not available.

\SetKwComment{Comment}{/* }{ */}
\SetKwInOut{Input}{Input}
\SetKwInOut{Output}{Output}
\RestyleAlgo{ruled}

\begin{algorithm}[hbt!]
\caption{Warped Multifidelity Gaussian Process, description of the steps involved in the derivation of WMFGP estimates of the mean $\mu_H(\mathbf{x^*})$.}
\label{algo3}
\Input{$D = [\mathbf{y_H}, \mathbf{y_L}]$}
\Output{WMFGP interpolation of the mean $\mu_H(\mathbf{x^*})$}
\For{$j$ in $D$}{
    \textbf{a.} Perform kernel density-based estimation and derivation of the bandwidth $h$\;
    \textbf{b.} Compute the kernel-based estimate of the CDF to derive $p_i$ probability levels based on the sample values $\mathbf{y}$ of the time-series, the chosen kernel and estimated $h$\;
    \textbf{c.} Compute the normal scores by inverting $\{\Phi^{-1}(p_i)\}_{i=1}^{N} = \mathbf{Ny}$\;
    \textbf{d.} Interpolate between the estimated CDF estimates ($p_i$) at the data points (either $\mathbf{y_H}$ or $\mathbf{y_L}$) to obtain a dense grid (4000 points) of $p_i$, at the query points z-scores having the following range $([\mathbf{y_{\text{min}}} - h, \mathbf{y_{\text{max}}} + h])$\;
    \textbf{e.} Generate a lookup table to link the actual data with their probability levels. The table contains a vector z-scores the corresponding interpolated $p_i$\;
}
\textbf{f.} Run the MFGP on the normal scores $\mathbf{Ny}_L$ and $\mathbf{Ny}_H$\;
\textbf{g.} Use the lookup table to backtransform the MFGP estimations\;
\textbf{h.} Return $\mu_H(\mathbf{x^*})$\;
\end{algorithm}

The lookup table contains two columns: the first one holds the query points (z-scores), while the second column holds the corresponding probability levels. These query points form a dense grid of values ranging from $[\mathbf{y_{\text{min}}} - h, \mathbf{y_{\text{max}}} + h]$.
Some of the probability levels ($p_i$) come from linear interpolations, while others are the actual values ($\mathbf{Ny}$) determined using the inverse of the standard normal distribution function ($\Phi^{-1}$). This means that the warping function can be represented as $\{\Phi^{-1}(p_i)\}_{i=1}^{N} = \mathbf{Ny}$. Linear interpolation is performed using z-values as query points. The actual data, either $\mathbf{y_H}$ or $\mathbf{y_L}$, serve as the independent variable, while the estimated cumulative density function values ($p_i$) act as the dependent variable. This interpolation establishes a connection between the actual data and their estimated cumulative density function values within a dense grid of z-values. After executing multifidelity models in the latent space, the estimates are expressed in terms of normal scores. These normal score estimates, denoted as $\mathbf{N\hat{y}}$, are then input into the normal cumulative distribution function to derive the estimated probability level ($\hat{p_i}$). For each estimated $\hat{p_i}$, we identify the nearest values $p_i$ in the second column of the lookup table and extract their indices. Subsequently, these indices are utilized with the ordered vector of z-scores, representing the first column in the lookup table, to extract predictions on the original scale. A concise description of the method is provided in Algorithm \ref{algo3}.

\section{3. Data}

\subsection{3.1 Data for the first simulation experiment}
The data for the first simulation experiment are based on the seasonal and trend decomposition of the wind speed reanalysis time-series downloaded from the Copernicus Climate Data Storage (CCDS), see reference ERA5 \cite{ERA5}. The CCDS offers a wide range of open access climate data, with in depth description and of the production process.
The single level\footnote{see ERA5 website for more information about single level dataset.} dataset offers a temporal coverage from 1940, with a hourly temporal resolution and spatial resolution of 0.25° x 0.25°, with data updated with a latency of 5 days. We chose offshore wind speed data from a site designated for a future wind farm. While the exact location is not critical for our current project, we made this decision to uphold the principle of replicating wind speed measurements in areas of potential significance. The vectorized wind speed components $u$ and $v$ are obtained at an altitude of 100 meters, and the wind speed is reconstructed using the Pythagorean theorem, i.e., \textit{wind speed} $= \sqrt{u^2 + v^2}$. A seasonal and trend decomposition using LOESS \cite{cleveland1990stl} is applied to extract the structural components (denoted as $T$ in Equation \ref{SimTS}) of the downloaded time-series. These components serve as the basis for the simulation discussed in Section \ref{firstsim}.

\subsection{3.2 Data for the application and second simulation experiment}
According to Italian and European legislation, the monitoring and protection of the environment (e.g., quantification of air and water pollution levels, measurement of meteorological phenomena, checks and inspections on the environmental impacts of enterprises, and scientific support to national and local institutions) is entrusted to the Regional Environmental Protection Agencies (ARPAs) that are members of the National System for Environmental Protection (SNPA)\footnote{It is an Italian acronym  of Sistema Nazionale di Protezione Ambientale}. Each ARPA is in charge of gathering air quality information within the area of responsibility. They also collect data on various environmental variables such as wind speed, temperature, and humidity. Data are publicly available either using the agency's data portal (\cite{ARPA}ARPA Data Portal), the Regione Lombardia open data portal (\cite{LombardyWeather}Lombardy Region Weather Station) or by means of specialized oper source software, such as \textit{ARPALData} (\cite{maranzano2023ARPAldata}) and EEAaq (\cite{EEAaq}). Here the \textit{ARPALData} package was used to retrieve the weather measurements used in both the simulation design and the application. Measurements span from 1st January to 31st December, 2022. 

\begin{figure}
        \centering
        \includegraphics[width=0.8\linewidth]{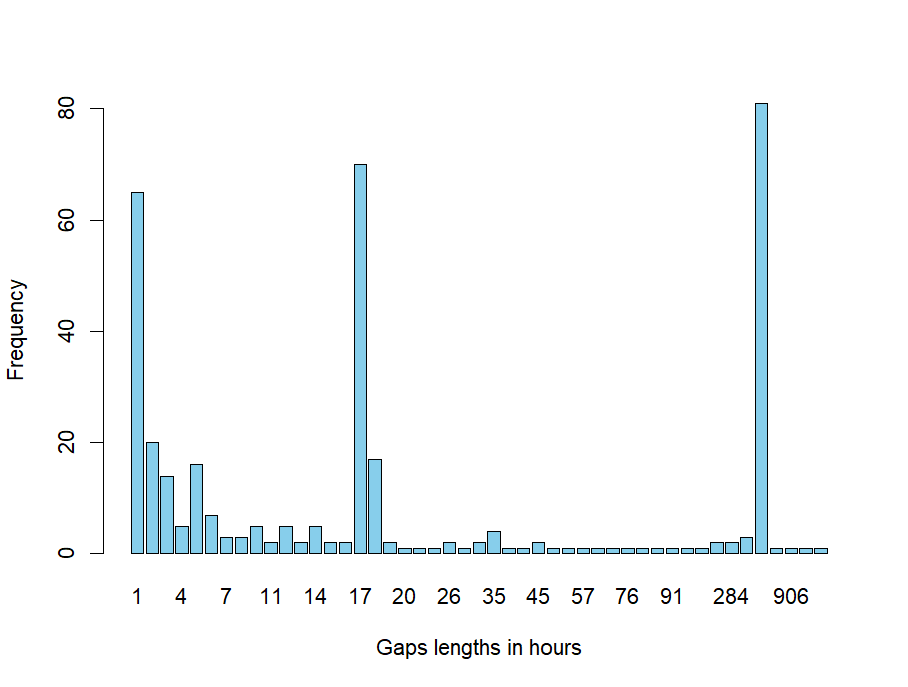}
        \caption{Frequency and duration of missing data sequences for all monitoring stations of Lombardy in the 2022. For a map of stations position see Figure \ref{fig:clustering}.}\label{MissingValues}
   
\end{figure}

The Lombardy network comprises more than 120 ground sites monitoring wind speed and direction, some of them active since the 90's and others activated only in the last few years. Therefore, a subset of 94 monitoring stations for wind speed activated before January 1st, 2015 and scattered throughout the region is investigated in this paper. Figure \ref{fig:clustering} displays the illustration of the station position and territory characteristics. By default, weather stations collect data at 10-minute intervals, however, in both the simulation experiments and the application, we considered the hourly average wind speed as we interested in recovering general patterns, however, the dataset allows for higher resolution studies. Figure \ref{MissingValues} illustrates the frequency and duration in hours of missing data sequences in the Lombardy region throughout 2022. The histograms reveal numerous gaps with duration of less than 15 hours, which can be addressed using standard interpolation methods. Gaps ranging from 15 to 192 hours (between one day and more than one week) are of particular interest in this application, as they are difficult to fill using simpler interpolation techniques. Longer gaps exceeding 192 hours are not considered, as they likely indicate structural malfunctions rather than structural missing data and it would be unwise to address these only using statistical methodologies.

\section{4 Simulation studies}
In this section, we introduce two experimental designs to evaluate the performance of the warped multifidelity Gaussian process in processing skewed data. These experiments illuminate the inherent challenges and effective strategies for reconstructing missing sequences inside a HF time-series. The first experiment assesses how well a series\footnote{See Section \ref{firstsim} for the list of the models.} of models performs when faced with random patterns of missing data in high-fidelity time-series. This scenario is standard in multifidelity modelling, where high-fidelity (HF) data are randomly sampled with equal probability. In this case, our goal is to understand how different models handle the task of recovering missing skewed HF information when HF data is sparsely sampled. We refer to this experimental design as \textit{randomized missingness} since the sampling frequency of HF data is very low (10$\%$) and occurs randomly. The second experiment addresses the issue of structural missingness, which involves the presence of long sequences of missing data in a time-series. In this latest experiment, following \cite{SERRA2023}, rather than simulating random locations and temporal patterns, we employ actual series from the ARPA Lombardia network in a Monte Carlo experiment, where the real locations are randomly sampled. The map of the monitoring stations and of the geography of Lombardy, shown in Figure \ref{fig:clustering}, evidences that in northern part of the region there is prevalence of mountains, with a higher spatial concentration of monitoring stations, while in the southern part, where the plains and  cities prevail, the stations are more sparsely scattered. Due to the absence of physical obstacles, the correlations of the time-series recorded in the southern region will be higher, potentially bringing a more efficacious recovery of the missing sequence. 
In both experiments, we employ simulations mimicking real-world data scenarios to analyze the model's performance, focusing on the challenges posed by skewed data. Our data fusion methodology, is applied to the data relying on a single assumption of linear correlation\footnote{In our application the correlation is always positive.} between datasets. Among other benefits, our methodology is able to impute long missing sequences (up to a week of data with hourly resolutions). Moreover, it can manage different varying seasonal, sub-seasonal, and cyclic components automatically, without ad hoc study for each fused time-series.

\subsection{4.1 First simulation design: randomized missingness.} \label{firstsim}
This experiment assesses how various models perform under changing conditions when recovering randomized patterns of missing information in high-fidelity time-series data. For this simulation experiment, we employed a data generation procedure to replicate wind speed data at specific locations of interest. In this context, we generated two wind speed time-series—one with limited reliability but a higher number of data points in time and the other with greater reliability but fewer data points. Although our simulation mimics wind speed data, the experiment can be generalized to any skewed data sources characterized by a randomized data missingness structure. The experiment involves the comparison of five models across different skewness levels. The selected models include a standard Gaussian process (GP), a warped Gaussian process (WGP), a classic multifidelity (MFGP) model, a multifidelity model integrated with a simple Box-Cox class transformation (BCMF), and finally, a warped multifidelity (WMFGP) model. We conducted tests in the following dimensions: 
\begin{enumerate}
    \item Randomized missingness: each experiment simulates a different missing data pattern, with a total of 200 replications.
    \item Level of skewness: time-series with varying degrees of skewness were generated, where skewness represents a parameter describing the deviation from normality. 
    %\item Level of Noise: We compared the models under varying noise conditions. 
\end{enumerate}
The time-series were generated following this structure:
\begin{align}\label{SimTS}
\begin{split}
\mathbf{y_L} = T + w_L \\
\mathbf{y_H} = T + w_H.
\end{split}
\end{align}

In this context, \textit{$T$} represents the true time-series we intend to recover While \textit{$w_L$} and \textit{$w_H$} denote random errors generated from skewed distributions. Initially, we generated \textit{$w_H$} from a closed-skew normal distribution parametrized as follows: $CSN(\mu, \sigma_1, \gamma, \nu, \delta)$, where $\mu$ represents the location parameter, $\sigma_1$ is the scale parameter, $\gamma$ is the skewness parameter, $\nu$ is the shape parameter, and $\delta$ is the truncation parameter. We also generated data from another distribution commonly used for wind speed data, allowing for more extreme skewness scenarios: the Weibull distribution with shape and scale parameters. We simulated two scenarios for both distributions, labelled as \textit{high-skewness} and \textit{low skewness}. The parameters for both distributions for these high and low fidelity errors (\textit{$w_L$} and \textit{$w_H$}) are detailed in table \ref{table CSN} and for the Weibull table \ref{tab:high_skew}.

\begin{table}
%% Caption MUST come immediately after \begin{table}
\caption{\label{table CSN}Parameters of the CSN distribution for the different skewness scenario.}
\centering
\begin{tabular}{l|r|r|r}
& \textbf{Low skewness } & \textbf{High skewness}  \\
 \textbf{Data} &  $w_H$ , $w_L$ & $w_H$ , $w_L$\\ 
\hline
$\mu$ & -0.25,-0.5  & -0.25,-0.5\\
\hline
$\sigma_1$& 0.04,0.8 & 0.8,2.4  \\
\hline
$\gamma$ & 4,4 & 50,50 \\
\hline
$\nu$ &2 & 2\\
\hline
$\delta$&3 &  3
\end{tabular}
\end{table}

\begin{table}
 \caption{ \label{tab:high_skew} Summary of statistical measures and generating parameters of the errors generated from the Weibull distribution, for different skewness scenario.} 
    \centering
  
        \centering
        \begin{tabular}{l|l|l|l|l|l|l}
     High Skewness & \textbf{Error} & \textbf{Scale} & \textbf{Shape} & \textbf{Mean} & \textbf{SD} & \textbf{Skewness} \\
            \hline
            &\textbf{$w_L$} & 2 & .8 & 1.3 & 2.8 & 2.8 \\
            &\textbf{$w_H$} & 0.5 & .8 & 0 & 0.72 & 2.9 \\

 % Add vertical space between the tables
    
     Low skewness & &  & & &  \\
            \hline
            &\textbf{$w_L$} & 2 & 2.3 & 1.18 & 0.82 & 0.46 \\
            &\textbf{$w_H$} & 0.5 & 2 & 0 & 0.23 & 0.66 \\
        \end{tabular}
      
\end{table}  

As we simulated missingness in the  HF data, we conducted model comparisons on the test set using the following performance metrics: Mean Absolute Error (MAE), Bias, and Variance (Var)\footnote{For the purposes of this paper, we focused only on the MAE and its variability.}. For consistency, we maintained the HF sample size at a fixed $10\%$ of the low-fidelity data. It's important to highlight that in employing the multifidelity approach, the ratio $\frac{N_H}{N_L}$ plays a pivotal role as it validates the utility of multifidelity models. With a high ratio the probability of the multifidelity models to not yield superior outcomes compared to standard monofidelity models is increased. Specifically, our analysis reveals that multifidelity proves advantageous up to a ratio of $30\%$ to $35\%$ in this particular data configuration.
\begin{figure}
        \centering
        \includegraphics[width=1\linewidth]{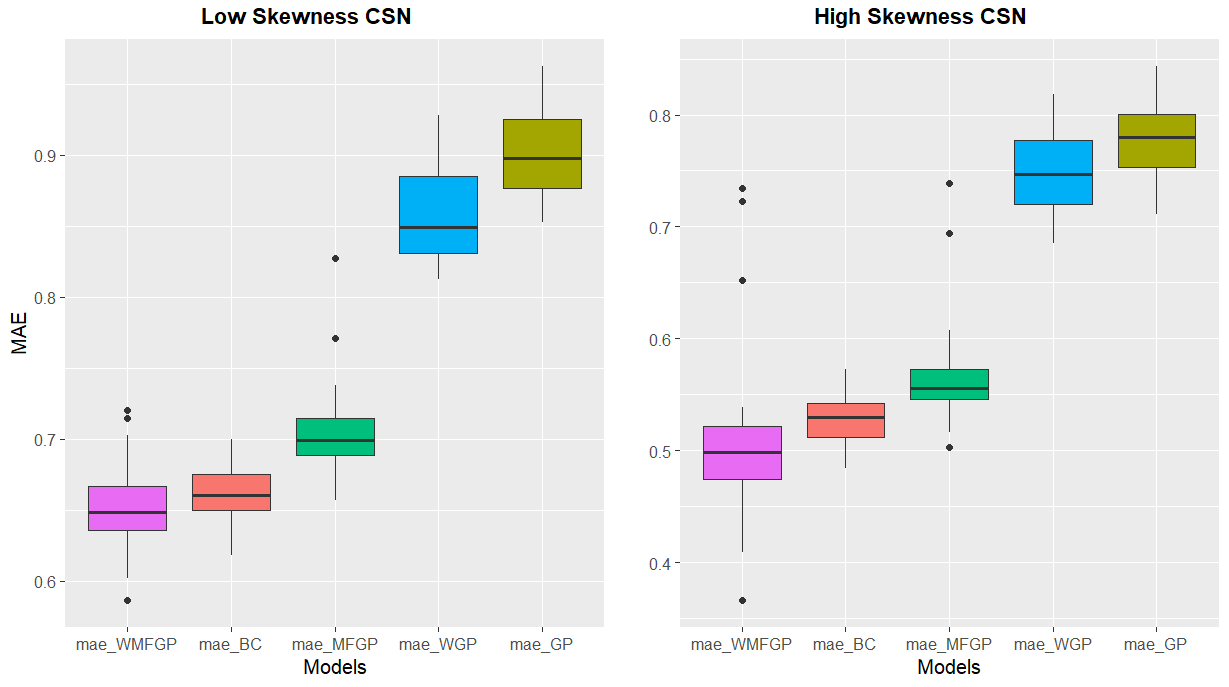}
        \caption{Boxplots of the 200 simulation experiment. On the $x$-axis the models, while on $y$-axix the MAE. Random errors generated from a closed skew normal distribution. Each box-plot refers to Mean Absolute Error in  the test set for each model. Note that WMFGP has the lowest performance in both scenarios.}
        \label{CSNexperiment}
\end{figure}
\begin{figure}
        \centering
        \includegraphics[width=1\linewidth]{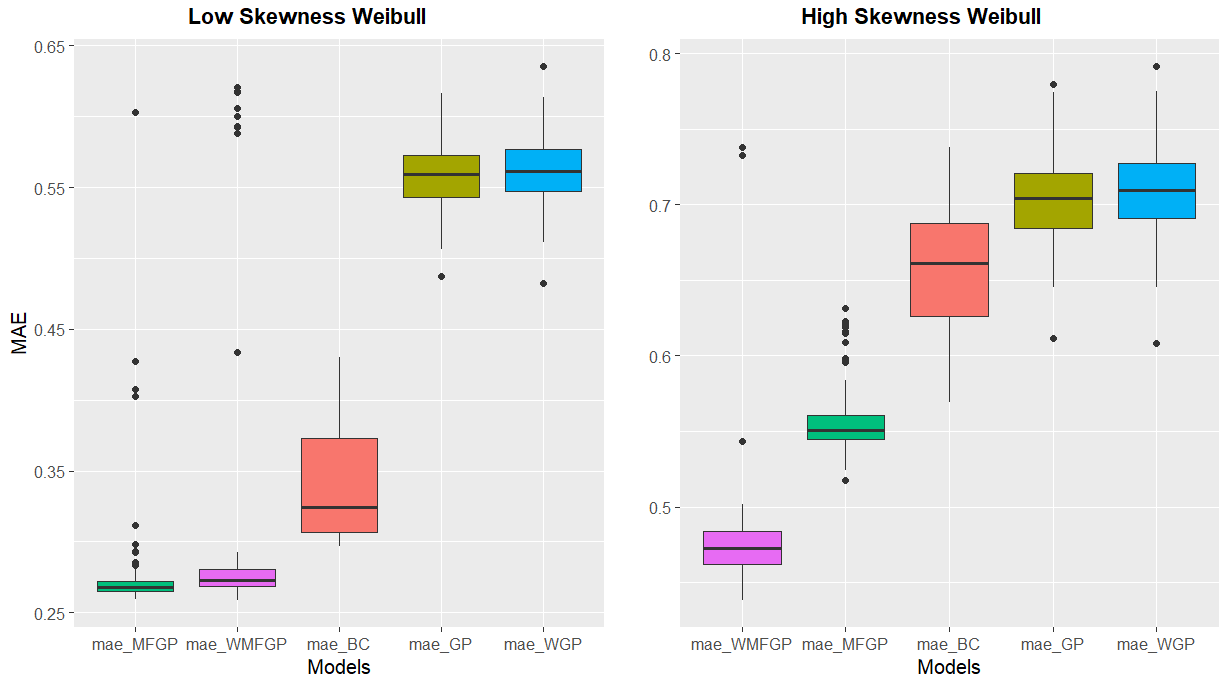}
        \caption{Boxplots from 200 simulation experiments are presented here. The models are listed on the $x$-axis, and the Mean Absolute Error (MAE) is displayed on the $y$-axis. Random errors in these simulations were generated from a Weibull distribution. Each boxplot represents the MAE on the test set for each model. In scenarios with high skewness, the Warped Multifidelity Gaussian Process (WMFGP) demonstrated superior performance. The performance of BCMF, highlighted in red, deteriorated as the underlying data became noisier and exhibited higher variance.}
        \label{Weibullexperiment1}
\end{figure}

In all four scenarios, the WMFGP model emerged as the top-performing model, as illustrated in the bottom panels of Figures \ref{CSNexperiment} and \ref{Weibullexperiment1}. It is noteworthy that, in general, the performance of the WMFGP model remained stable independently of the skewness level, while simpler multifidelity models face challenges. Occasionally, all multifidelity models (WMF, MF, BCMF) exhibited anomalous performances attributable to numerical instability, as evidenced by certain outlier data points in the boxplot. However, in synthesis, the WMFGP consistently outperforms the other models. The marginal improvement seen with the small skewness Weibull distribution under the standard multifidelity model is minor. The WMFGP resulted to be the best methods in 3 out 4 scenarios, which more than expected since the method is thought to provide an advantage only in high-skewness scenarios. The Box-Cox multifidelity model generally was a good method, when the error was generated from a closed skew-normal, while when the error was generated from Weibull, where overall the skewness condition were more extreme, it was not able to returns good performance. The latter confirmed the expectations: the Box-Cox method exhibits high sensitivity to data; when dealing with multiple datasets, the parametric transformation required to normalize each data source might vary significantly. This can inevitably affect the relationships between datasets in the latent space. Clearly, the more different the datasets are, the more likely it is that they will require different transformations to be normalized.

%The compared methods are not chosen at random. The Gaussian process regression models GP and WGP are included to verify at which sampling level of HF data makes sense to use a multifidelity model. The standard motivation behind using multi-fidelity models resides in the low availability of HF samples. Indeed, a significant enough sample of HF data would make the multifidelity models useless. The BCMF is introduced to prove that standard parametric transformation-based methods are inconvenient in a multi-fidelity scenario. Then, we included the classic MFGP as a benchmark for our WMFGP.

\subsection{4.2 Second simulation design: structural missingness} \label{struc_miss_ex}
We performed a second simulation experiment with the aim of proposing multifidelity methods for dealing with structural missingness. Structural missingness can be seen as the presence of long-missing sequences in a time-series. This issue is of particular concern since standard interpolation methods such as Gaussian process regression have very poor predictive power if, for example, the missing sequence is longer than the so-called range parameter or \textit{length scale} that controls the rate at which the covariance between two data points decreases as the distance between them increases. We focused on hourly resolutions, as we are concerned with recovering general patterns, we selected a sub-sample of 72 out of 94 stations from the network, to make sure that no real missing values where present in the experiment, with data referring to 2022. Then, we simulated missing sequences of the following lengths in hours:

\begin{equation} \label{missingSeq}
    \text{missing lengths} = [24, 48, 72, 96, 192]
\end{equation}

where 24 stands for 24 hours of missing values. We tested the capacity to recover the missing sequences using five methods: the GP, Surrogate, MFGP, WMFGP, and a simple imputation (SI). In this experiment, we excluded BCMF, since, as shown in the previous experiment, the method was never better than a standard MFGP method. However, we included a SI method as a benchmark for comparison. The simple imputation is a fast implementation method that assumes the values of the missing sequence to be an average of the closest observations, a moving average k-nearest neighbour. The wind speed data used in this experiment are generally skewed; therefore, our expectations are that the WMFGP will outperform MFGP. The surrogate performance represents the discrepancy we were to obtain if we replaced the missing sequence with the data observed in the nearby monitoring station, that we labelled as LF data source. In simple terms, surrogate is just $LF-HF$. The surrogate performance is a benchmark; we want the discrepancy between the estimates and the HF to be at least smaller than the differences between HF and LF data. Intuitively, once the surrogate resulted are available, we can measure the percentage error reduction of each method when compared to the original discrepancy between LF and HF.
Our strategy rely on the principle that monitoring stations that are in the same neighbourhood often present correlated information but different structural missingness. In this experiment, we assume the HF data to be those with a missing sequence, while the LF data comes from a nearby highly correlated station. 
The table \ref{Tabel_structuralMissing} summarizes the medians\footnote{The medians are chosen to limit the impact generated by the numerical instability of the algorithm.} of Mean Absolute Error (MAE) performance along with their respective standard deviations (sd) obtained from a series of 500 replication experiments. 
These experiments were conducted to assess the performance of five different imputation methods when dealing with simulated missing data sequences of varying lengths. Each row of the table corresponds to a specific missing data (ML) sequence length, denoted as ML:24, ML:48, ML:72, ML:96, and ML:196. The columns represent the different imputation methods used: GP, MFGP, WMFGP, Surrogate and Simple Imputation. 
The values in the table show the medians of MAE scores for each combination of imputation method and missing data sequence length, the median have been choose to limit the impact of numerical errors, so the importance of anomalous performance is reduced. Additionally, the standard deviations (sd) are provided in parentheses, giving an indication of the variability in the MAE scores across the 500 replications.

\begin{table*}[!t]
\caption{The table contains medians of MAE performances and their standard deviations for each simulated missing sequence. ML 
 stands for missing sequence length.}\label{Tabel_structuralMissing}
\tabcolsep=0pt
\centering
\begin{tabular*}{\textwidth}{@{\extracolsep{\fill}}l|c|c|c|c|c}
& GP & MFGP & WMFGP & Surrogate (abs(HF-LF)) & Simple Imputation \\
\midrule 
\textbf{ML:24(sd)} & 0.79 (0.71) & 0.55 (0.45) & 0.46 (0.51) & 0.75 (1.05) & 1.48 (1.36) \\
\hline
\textbf{ML:48 (sd)} & 0.77 (0.69) & 0.57 (0.52) & 0.49 (0.54) & 0.75 (1.05) & 1.62 (1.18) \\
\hline
\textbf{ML:72 (sd)} & 0.79 (0.59) & 0.52 (0.48) & 0.47 (0.49) & 0.74 (0.89) & 1.44 (1.16) \\
\hline
\textbf{ML:96 (sd)} & 0.80 (0.60) & 0.54 (0.44) & 0.48 (0.45) & 0.77 (0.85) & 1.43 (1.00) \\
\hline
\textbf{ML:196 (sd)} & 0.84 (0.40) & 0.51 (0.34) & 0.45 (0.34) & 0.71 (0.65) & 1.71 (0.89) \\
\bottomrule
\end{tabular*}
\end{table*}

%What is the overall trend ?
%What is the missing sequence length impact?
The trend for MFGP and WMFGP in the table is that the MAE tends to increase as the discrepancy between LF and HF (Surrogate LF) data increases. This suggests that the missing sequence lengths do not play a role in the performance of the methods. This is an interesting a result as in previous interpolation studies, based on simple Gaussian process, the gap length was associated with increased uncertainty (\cite{colombo2022quantifying} and \cite{fasso2020interpolation}). 
Considering the overall performance, WMFGP appears to be a strong candidate for imputing missing data, as it maintains relatively low MAE values across different missing sequence lengths. This result is generally expected as the wind speed data from the monitoring stations of Lombardy are generally quite skewed (from 0.3 to 2). The second best method is MFGP, confirming the using multiple data-sources is beneficial. The simple imputation method is not a good choice since such  high resolution of the data makes it difficult to understand what the smaller seasonal and cyclic components are. Figure \ref{fig:gap96} presents a simulated HF wind speed data reconstruction. In the top panel, we display the training data, which consists of LF data points shown as a blue dots, the target HF data represented by red square, and the interpolation area highlighted in yellow. Moving to the bottom panel, we showcase the recovery of HF information in the yellow area using two different approaches: MFGP, represented by the light blue dashed line, and WMFGP, shown as the magenta dashed line. The red dots on this panel represent the test data to recover. Notably, for this specific scenario, both the MFGP and WMFGP methods demonstrate their proficiency in approximating the missing information effectively, with a relatively better approximation performed by WMFGP. The plot also incorporates the $95\%$ prediction interval of the MFGP model, indicated by the shaded magenta area.  Interestingly, only one data point out of the 96 in the test area falls outside the prediction interval. The average discrepancy between the surrogate LF time-series and HF time-series ranged between 0.85-1.26 m/s in terms of wind speed. Meanwhile, the prediction of the WMFGP reduces the discrepancy to 0.45-0.65 m/s. These results constitute almost a 50$\%$ error reduction.

\begin{figure}
    \begin{subfigure}
    % {\textwidth}
        \centering
        \includegraphics[width=0.50\linewidth]{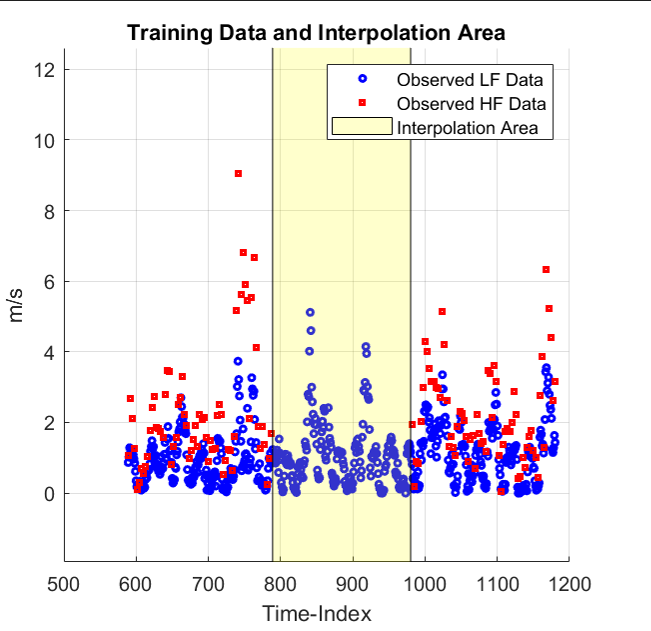}
    %   \caption{Training data for HF in red and LF in blue. The simulated test area is highlighted in yellow. The vertical axis represents wind speed in meters per second (m/s), while the horizontal axis denotes the time index in hours from the first observation.}
    \end{subfigure}
    \begin{subfigure}
        \centering
        \includegraphics[width=0.50\linewidth]{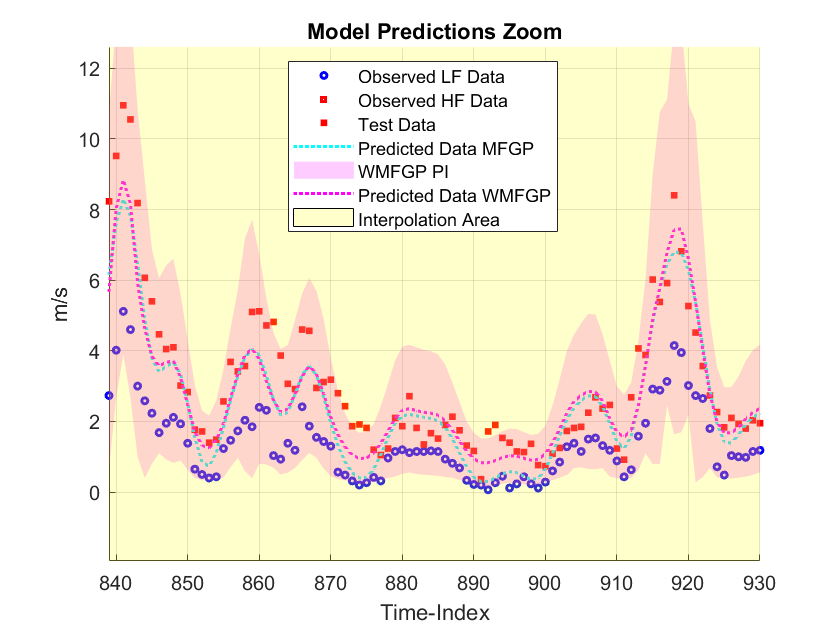}
        \caption{In the top panel, the training data for High-Fidelity (HF) simulations are depicted in red, while Low-Fidelity (LF) simulations are shown in blue. The simulated test area is highlighted in yellow. This bottom illustration showcases the successful recovery of 96 HF wind speed data points, emphasizing the robust performance of multifidelity models. Notably, across various time frames, the MWFGP – represented by the dashed magenta line – closely matches the simulated missing data (depicted as red squares) better than the MFGP, indicated by the dashed blue line. The vertical axis represents wind speed in meters per second (m/s), while the horizontal axis denotes the time index in hours from the initial observation.}
    \end{subfigure}
\end{figure} \label{fig:gap96}

\subsection{4.3 Uncertainty Discussion}
 Discussing the uncertainty is of paramount importance for informed decision-making, risk management and resource allocation. In the structural missingness simulation study, we calculated both the prediction intervals and their corresponding coverage probability \footnote{The coverage probability serves as a measure of the prediction interval's accuracy. In essence, it indicates the proportion of times, within a repeated or hypothetical series of experiments or predictions, that the prediction interval successfully encompasses the true value of the variable of interest.}. This measure is typically expressed as a percentage or a probability value. During 500 simulation experiments, we determined that the coverage probability of the prediction interval, established with a 95$\%$ confidence level for the MFGP, was 91$\%$. Similarly, for the WMFGP, it was found to be 90$\%$. This suggests that, on the whole, our prediction interval computations exhibit a high degree of accuracy. There are several reasons why a prediction interval may not achieve perfect alignment with 95$\%$ confidence level. For instance, in some experiments, the multifidelity class may not be the most appropriate choice, while numerical instability might come into play in others. However, despite these potential sources of variability, the coverage probability remains satisfactory, given the inherent complexities and possible challenges , section \ref{Application}.

\section{5. Application} \label{Application}
In the previous sections we stated that the ARPA monitoring networks often contain numerous missing values with different gap lengths. Therefore, we utilize the WMFGP and MFGP methods to recover these gaps in the sequences. Our approach leverages the natural temporal correlation among time-series observed at nearby locations, resulting in a highly computational efficiency strategy and of easy implementation. By excluding the spatial dimension from the multifidelity part, we might incur endogeneity issues \cite{le2021endogeneity}, as external factors like orography, geography, and land cover can strongly influence the model's performance. To address this concern, we conducted a clustering experiment based on the latitude, longitude, and altitude of the stations.

\subsection{5.1 Empirical strategy}
This procedure on one hand aims to identify monitoring stations that can serve as surrogate time-series data (LF); on the other to fill the gaps a target high-fidelity (HF) time-series data. The assumption is that nearby stations are likely to have similar information \cite{zhu2022third}, and therefore, correlated data can be used in a multifidelity context. We follow these steps to discover clusters of stations:

\begin{enumerate}
    \item Constrained k-means: We utilize constrained k-means to identify clusters based on spatial information about the monitoring stations, such as latitude, longitude, and altitude. We also set constraints on the maximum and minimum cluster sizes. More details about the specific functioning of this method are available in \cite{bradley2000constrained}. Fixing the number of clusters is crucial because, for our purposes, it makes little sense to have clusters containing, for example, 20 stations, even if they are well-defined.
    \item Randomized pairing: We select a station with gaps and pair it with another randomly selected station within the same cluster.
    \item The stations with gaps constitute our HF dataset, while the paired stations serve as the LF dataset. We then run multifidelity models using these two datasets.
\end{enumerate}

\begin{figure}
    \centering
\includegraphics[width=1\linewidth]{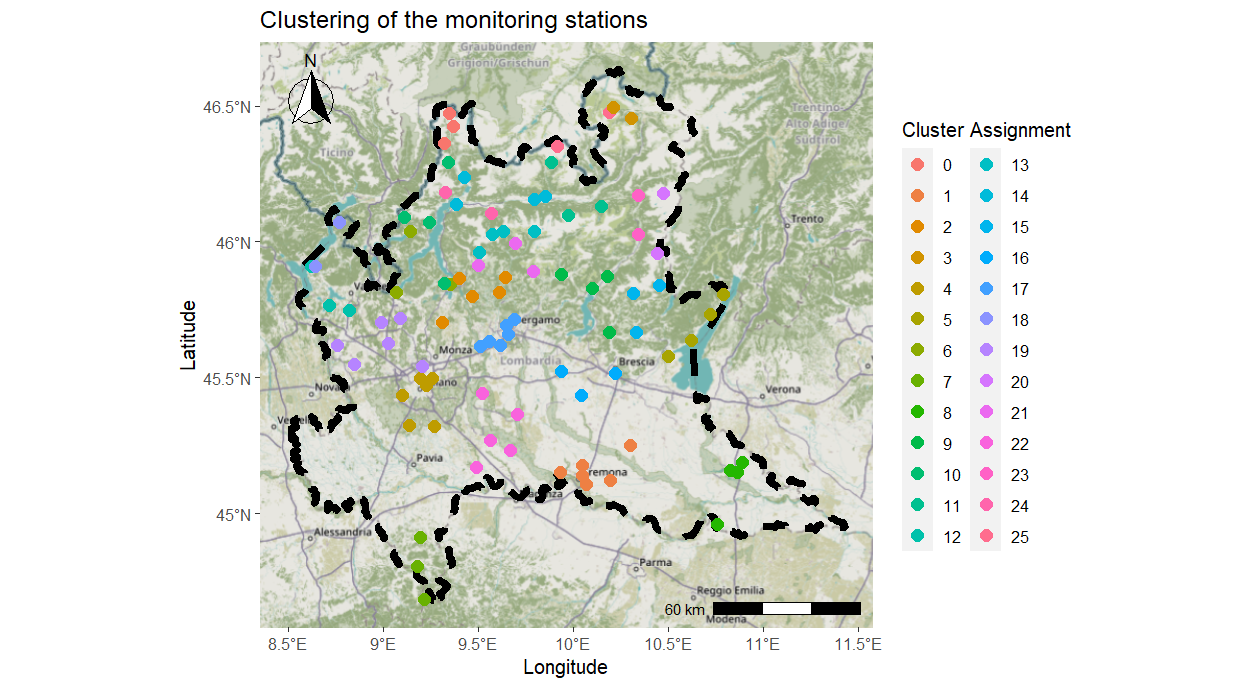}
    \caption{Depiction of the monitoring station clustering in two dimension (longitude and latitude).}
    \label{fig:clustering}
\end{figure}

The results of the constrained k-means are shown in Figure \ref{fig:clustering}. The plot is shown in two dimensions for clarity; however, we also used the altitude for clustering purposes. The altitudes play a crucial role in defining the clusters as in the north part of the region, the presence of the mountains hinders the cluster aggregation; in the same way, stations at similar latitudes and longitudes but different altitudes might have very different wind speeds.

%\begin{Figure}
%  \centering
%  \begin{subFigure}{0.48\textwidth}
%    \includegraphics[width=\linewidth]%{BestClusters.png}
%    \caption{Depiction of the monitoring %station clustering in two dimension %(longitude and latitude).}
%    \label{fig:clust2d}
%  \end{subFigure}
%  \begin{subFigure}{0.48\textwidth}
%    \includegraphics[width=\linewidth]%{3dPlot_oftheClusters.png}
%    \caption{Depiction of the monitoring %station station clustering in 3 dimensions %(longitude, latitude, and altitude).}
%    \label{fig:clus3d}
%  \end{subFigure}
%  \caption{Clustering of the ARPA monitoring %stations.}
%  \label{fig:clustering}
%\end{Figure}

\begin{figure}
    \centering
    \includegraphics[width=1\linewidth]{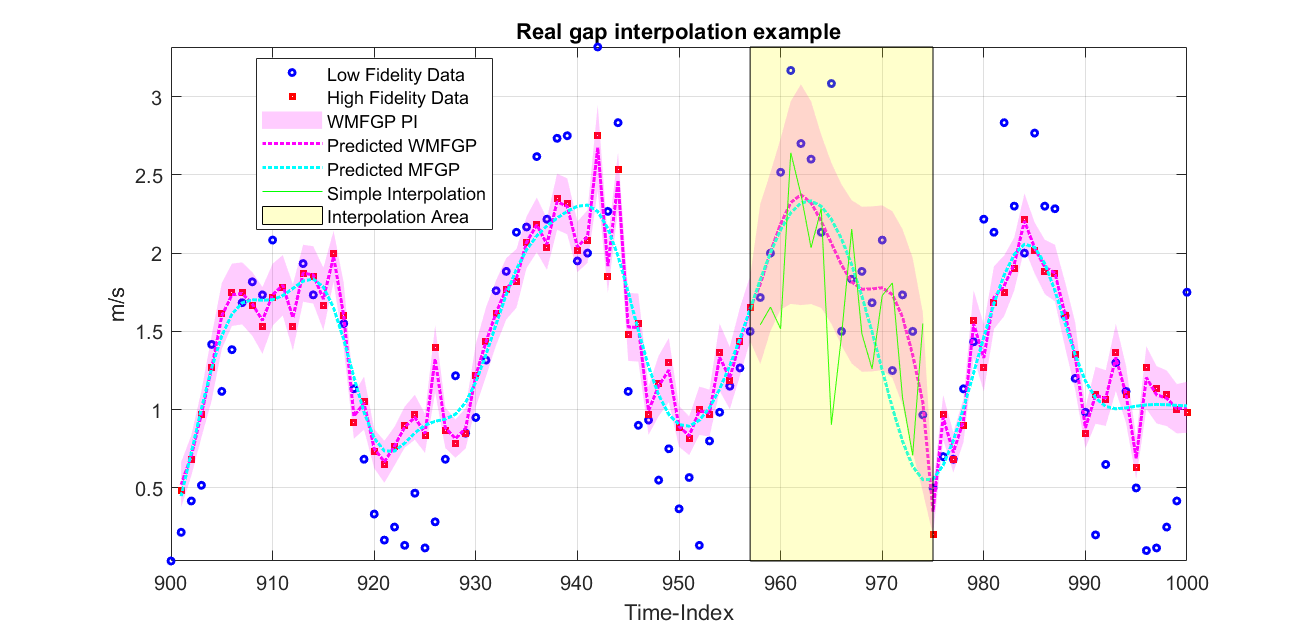}
    \caption{Depiction of the prediction performed by MFGP (light blue), the WMFGP (magenta) and the SI (green) for the Veddasca Monte Cadrigna station. Note that in the interpolation area in yellow, SI returns and  unlikely recovery of the HF information, while both the WMFGP and MFGP seem to return a plausible pattern.}
    \label{fig:RealImputation}
\end{figure}

We obtained 26 clusters by optimizing standard metrics, such as the elbow plot and average silhouette. This experiment resulted in an average cluster size of 3.6, with the largest cluster containing six elements and the smallest containing three elements. Cluster sizes ranging between 3 and 6 are convenient for two main reasons. Firstly, if the same missing sequence is present in two nearby stations, we can select the data from the third station of the clusters. This scenario, where the same gap appears in three nearby stations, is implausible in our dataset. Secondly, avoiding larger groups reduces the probability of associating stations influenced by different phenomena. Figure \ref{fig:RealImputation} illustrates the imputations of 17 observation gaps for the Veddasca Monte Cadrigna station. Based on the simulation experiment, we know that among stations sharing similar altitude, longitude, latitude, and correlation between LF and HF, the MAE of WMFGP and MFGP was 30\% lower that the MAE of the Surrogate. More importantly, the WMFGP had a MAE 13$\%$ lower than the MFGP. The latter is aligned with what is shown in Figure \ref{fig:RealImputation}, where the seasonality is correctly recovered by both the multifidelity models, but with a higher adapatability of WMFGP(magenta line), when compared to the MFGP (light blue). The green line represents a simple interpolation method, which, in this specific scenario, provides a poor recovery of the HF signal.
%It is important to note that other simpler methods like STL decomposition may effectively decompose the signal into its structural components (seasonality and cycle); however, each station may have a different cycle and seasonality, or more precisely harmonics with different a non stationary amplitude and wave periods necessitating an ad hoc study for each missing sequence. Another limitation is that for longer imputations, such as a week of missing data, the STL strategy may be less effective for prediction tasks, especially when no evident cyclic or seasonal pattern is present, as seen in \ref{fig:gap96}, where the harmonics of the signal to recover have different amplitudes. 
The data imputation remains independent of the presence of different structural components, in different time-series. Meaning that if each sub-experiment involves time-series having different seasonal and cyclic components, it is not necessary to perform an ad hoc study to recover the right harmonic. In other words,
the advantage of MFGP data imputations lies in the possibility of automating the data imputation process by simply identifying a correlated time-series with the one of interest.  

\section{6. Conclusion}
Integrating multiple data sources can effectively leverage the relative abundance of low-quality data with the relative scarcity of high-quality data. Many data fusion algorithms based on Gaussian processes have been developed to model various linear and non-linear relationships between different datasets. However, little or no work has been done to account for the presence of non-normal properties in the datasets to be integrated. This latter aspect is an explicit limitation, considering that GP models assume independent, identical, normal errors, which could lead the model to have poor predictive performance. Moreover, the standard strategies for accounting for deviations from normality, such as parametric transformations, might not be straightforward to implement in a data fusion application where each dataset might require a different transformation, violating the principles of quantile invariance.

In this work, we proposed an extension of the autoregressive multifidelity Gaussian process, a standard method for resolving data fusion problems with Gaussian process, based on a non-parametric warping strategy. This extension helps deal with multifidelity applications where the discrepancy data exhibit skewness, a common situation when at least one of the different data sources presents skewed data. Since the model preserves the quantile invariance, it can be effectively applied to multiple data sources. The algorithm models the relationships between high-fidelity and low-fidelity observations in a latent space, where, owing to the warping, the discrepancy data follow a normal distribution, thereby improving the prediction of high-fidelity data. We showcase the efficacy of our model in modelling multiple data sources through an extended simulation experiment in which we controlled for the high-fidelity sample size and skewness. The model can be used to fill long missing sequences of environmental time-series. In this paper wind speed data have been filled, however, the model has the potential to fuse and fill data coming from any skewed time-series, such as time-series regarding pollutant concentrations or rain occurrence.
The wind speed data of the ARPA Lombardia monitoring network represented our motivating application as, due to both maintenance and adverse weather conditions, gaps of various lengths are often found in the measured time series. These gaps can cause problems when it comes to studying specific weather phenomena or when trying to assess the effect of wind speed on air quality. We, therefore, created a second simulation experiment to test the efficacy of multifidelity models to fill these long missing sequences. Our new method performs better than other tested methods and, particularly, better than the standard MFGP since the wind-speed data of Lombardy presented a considerable skewness. A nice feature of this method is that it can be automatized, while other imputation methods would require an ad hoc study for each time-series. This is an appealing characteristic when dealing with big environmental datasets. Another interesting result is that the multifidelity models seem to be independent of gap dimension see \ref{Tabel_structuralMissing}, suggesting that they could be used for imputing very long missing sequences. One limitation of the method might be related to the number of observations necessary to learn the warping. We observed that, generally, good warping transformations are learned when there are more than 1000 data points for each data source, which might not always be the case in data fusion applications. The simulation study and the application developed at this point are solely based on the temporal dimension (and correlation). This can be seen both as an advantage and a limitation. On one hand, the developed framework requires few computational resources, providing a ready-to-use strategy for different datasets.
On the other hand, it does not consider the additional strength that could be gained by incorporating the spatial dimension.  A spatial model could be useful in case of wind-farm resource assessment, where measurement taken with Lidar or anemometric technologies could be integrated with reanalysis data or it could be useful to map the wind speed in areas of Lombardy network that are not covered by any monitoring station. However, including the spatial dimension will necessarily increase the algorithm's computational complexity, requiring the implementation of suitable approximation methods for the inversion of the $\mathbf{K}$ variance-covariance matrix. Implementing a suitable approximation strategy in a multifidelity context is not trivial, as different points have different importance. We have reserved this line of research for a future study.

The strategy that we propose can be easily put into practise by practitioners and governmental agency that deal with missing sequence from a network. It is efficient, as it requires minimal computational resources, and it can be applied to any other datasets having similar characteristics to the ARPA Lombardia case study.

%\bmhead{Acknowledgements}

%Acknowledgements are not compulsory. Where included they should be brief. Grant or contribution numbers may be acknowledged.

\section*{Declarations}
\subsection*{Conflict of interest}
There is no conflict of interest to declare.
\subsection*{Data availability}
Data used in the paper are public (source Eurostat). Since we use only public data, no Special Permission is need to use copyrighted material from other sources (including the Internet). For reproducibility purposes, all scripts and the data are available at the following GitHub folder \href{https://github.com/PaoloMaranzano/RC_PM_RM_SCSAR_AgroConcentration.git}{https://github.com/PaoloMaranzano/RC\_PM\_RM\_SCSAR\_AgroConcentration.git}.

%\newpage

\bibliographystyle{unsrt}  
\bibliography{reference.bib}

\newpage

\appendix

\section{Appendix: extended results} \label{appendix1}

\begin{figure}[htbp]
    \centering
    \includegraphics[width=0.65\linewidth]{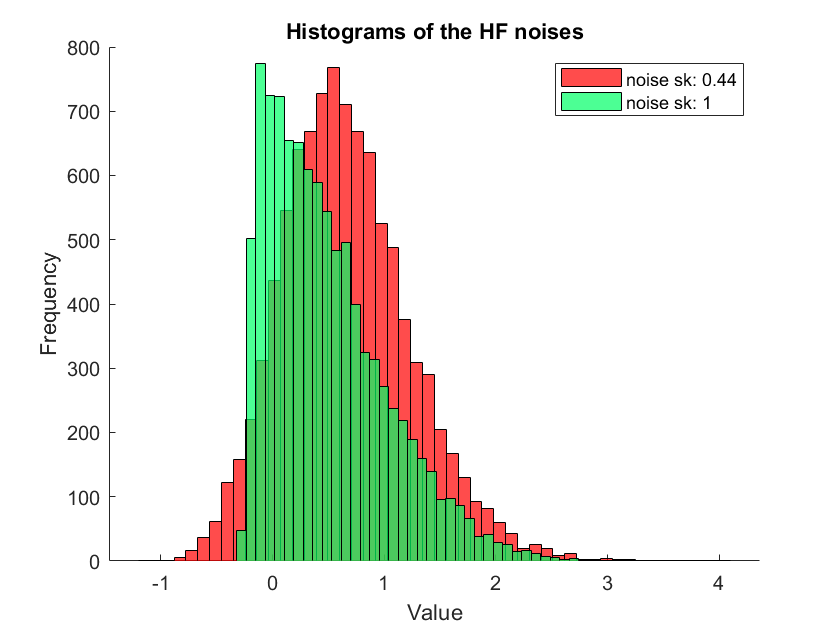}
    \caption{Histograms of HF data noise for different skewness scenarios. Noise values from a CSN distribution. $w_H$ depicted in red for low skewness, while $w_H$ in green for high skewness.}
    \label{NoiseCsn}
\end{figure}

\begin{figure}[htbp]
   \centering
   \includegraphics[width=0.65\linewidth]{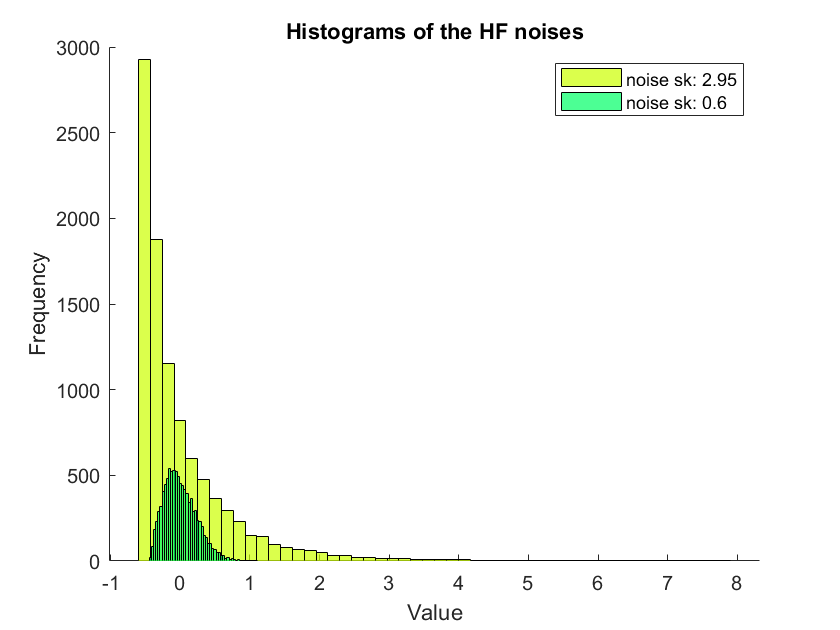}
   \caption{Histograms of HF data noise for different skewness scenarios. Noise values from a Weibull distribution. In this case, $w_H$ depicted in green for low skewness, while $w_H$ in yellow for high skewness. Note the increased skewness of both noises and the increased variance compared to Figure \ref{NoiseCsn}.}
\end{figure}

\newpage
\section{Limitations of the MFGP model for skew data.}
To understand the limitation of the MFGP, we have to understand the skewness of the discrepancies between two random variables. To determine the skewness of the difference of two random variables $aX - bY$, where $X$ and $Y$ are positively skewed and positively correlated, we can use properties of skewness and correlation. Let's denote the skewness of $X$ as $\text{Skew}(X)$, the skewness of $Y$ as $\text{Skew}(Y)$, and the correlation between $X$ and $Y$ as $\rho_{XY}$. The skewness of a linear combination of random variables can be expressed as follows:
\[
\text{Skew}(aX - bY) = \frac{E[(aX - bY)^3]}{\left(E[(aX - bY)^2]\right)^{3/2}}
\]
Given that $X$ and $Y$ are positively correlated, we have $\rho_{XY} > 0$. Now, we can calculate the third central moment $E[(aX - bY)^3]$ and the second central moment $E[(aX - bY)^2]$.

\[
E[(aX - bY)^3] = E[a^3X^3 - 3a^2bXY^2 + 3ab^2X^2Y - b^3Y^3]
\]
\[
= a^3E[X^3] - 3a^2bE[XY^2] + 3ab^2E[X^2Y] - b^3E[Y^3]
\]
\[
E[(aX - bY)^2] = E[a^2X^2 - 2abXY + b^2Y^2]
\]
\[
= a^2E[X^2] - 2abE[XY] + b^2E[Y^2]
\]
Given that $X$ and $Y$ are positively skewed, $\text{Skew}(X) > 0$ and $\text{Skew}(Y) > 0$, the third central moments $E[X^3]$ and $E[Y^3]$ are positive. Using the properties of expectation and correlation, we can determine the signs of the terms involved:

\begin{enumerate}
    \item $E[XY^2]$ and $E[X^2Y]$ will both be positive due to the positive skewness of $X$ and $Y$ and the positive correlation between them.
    \item $E[XY]$ will be positive due to the positive correlation between $X$ and $Y$.
\end{enumerate}

Therefore, all terms in the numerator of $\text{Skew}(aX - bY)$ are positive, and all terms in the denominator are positive as well. This implies that $\text{Skew}(aX - bY) > 0$. Thus, if $X$ and $Y$ are positively skewed and positively correlated, the difference $aX - bY$ will also be positively skewed.
 MFGP depends on two independent Gaussian processes: the discrepancy process $\delta(\cdot)$ and the low-fidelity process $u_L(\cdot)$. Both of these processes come with the standard Gaussian process assumptions, including the assumption that both $\epsilon_L$ and $\epsilon_{\delta}$ are independent and identically distributed normal errors. However, these assumptions might not hold in the presence of skewed data. More precisely, suppose that $\mathbf{x}$ is a generic vector of input locations. If both $u_H(\mathbf{x})$ and $u_L(\mathbf{x})$ are skewed, since $\rho$ is positive, their discrepancies will also be skewed. Rearranging the terms of equation \ref{Eq2}, we obtain:
\begin{equation}\label{eq9}
   \delta(\mathbf{x}) + \epsilon_{\delta}(\mathbf{x}) =  u_H(\mathbf{x})- \rho u_L(\mathbf{x}) ;
\end{equation}
This leads the first term of equation \ref{eq9}, which is standard Gaussian process, to model the skewed data resulting from   $u_H(\mathbf{x})- u_L(\mathbf{x})$, where $\rho$ is disregarded as it is a positive constant.  Since the normality assumption on $\epsilon_{\delta}$, we might incur in a imprecise high-fidelity process estimation, due to the inappropriate use of the discrepancy process.

\end{document}